\documentclass[12pt]{article} 
\pdfoutput=1

\usepackage[utf8]{inputenc}
\usepackage[T1]{fontenc} 
\usepackage{lmodern}
\usepackage[english]{babel}           
\usepackage{cancel}
\usepackage[pdftex]{graphicx}
\usepackage{epsfig}
\usepackage{graphicx}
\usepackage{comment}
\usepackage{latexsym}
\usepackage{hyperref}
\usepackage{amsmath}
\usepackage{mathrsfs}
\usepackage[usenames, dvipsnames]{color}
\usepackage{amsbsy}
\usepackage{amssymb}
\usepackage{amsthm}
\usepackage{amsfonts}
\usepackage{cite}
\usepackage{enumitem}
\usepackage{xcolor}
\usepackage{color}
\usepackage{mathtools}
\usepackage{caption} 
\usepackage{subcaption} 
\usepackage{euscript} 
\usepackage{float}
\usepackage{diagbox}
\usepackage[title]{appendix}
\usepackage{tabularx}

\def\be{\begin{equation}}
\def\ee{\end{equation}}
\def\bea{\begin{eqnarray}}
\def\eea{\end{eqnarray}}

\newcommand{\Vol}{{\rm Vol}}
\newcommand{\Diff}{{\rm Diff}}
\newcommand{\Id}{{\rm Id}}
\renewcommand\d{{\rm d}}

\newcommand{\bell}{\bar \ell}
\newcommand{\ba}{\bar a}
\newcommand{\bq}{\bar q}

\newcommand{\N}{{\cal N}}
\newcommand{\D}{{\cal D}}

\newcommand{\K}{{\cal K}}
\newcommand{\C}{{\cal C}}
\newcommand{\E}{{\cal E}}

\renewcommand{\S}{{\cal S}}

\newcommand{\wt}{\widetilde}

\newcommand{\dis}{\displaystyle}
\renewcommand{\thefootnote}{\fnsymbol{footnote}}

\newcommand{\Refe}[1]{Ref.~\cite{#1}}

\newcommand{\R}{\mathbb{R}}
\newcommand{\Z}{\mathbb{Z}}
\renewcommand{\natural}{\mathbb{N}}

\renewcommand{\O}{{\cal O}}
\newcommand{\cst}{{\rm cst.}}

\newcommand{\eg}{{\em e.g.} }

\newcommand{\where}{\mbox{where}}

\newcommand{\when}{\mbox{when}}
\renewcommand{\and}{\mbox{and}}

\newcommand{\bm}{\boldmath} 

\renewcommand{\theequation}{\arabic{section}.\arabic{equation}}

\catcode`\@=11
\def\marginnote#1{}
\newcount\hour
\newcount\minute
\newtoks\amorpm
\hour=\time\divide\hour by60 \minute=\time{\multiply\hour by60
\global\advance\minute by-\hour}
\edef\standardtime{{\ifnum\hour<12 \global\amorpm={am}%
        \else\global\amorpm={pm}\advance\hour by-12 \fi
        \ifnum\hour=0 \hour=12 \fi
        \number\hour:\ifnum\minute<10 0\fi\number\minute\the\amorpm}}
\edef\militarytime{\number\hour:\ifnum\minute<10 0\fi\number\minute}
\def\draftlabel#1{{\@bsphack\if@filesw {\let\thepage\relax
   \xdef\@gtempa{\write\@auxout{\string
      \newlabel{#1}{{\@currentlabel}{\thepage}}}}}\@gtempa
   \if@nobreak \ifvmode\nobreak\fi\fi\fi\@esphack}
        \gdef\@eqnlabel{#1}}
\def\@eqnlabel{}
\def\@vacuum{}
\def\draftmarginnote#1{\marginpar{\raggedright\scriptsize\tt#1}}
\def\draft{\oddsidemargin -.2truein
        \def\@oddfoot{\sl preliminary draft \hfil
        \rm\thepage\hfil\sl\today\quad\militarytime}
        \let\@evenfoot\@oddfoot \overfullrule 3pt
        \let\label=\draftlabel
        \let\marginnote=\draftmarginnote
   \def\@eqnnum{(\theequation)\rlap{\kern\marginparsep\tt\@eqnlabel}%
\global\let\@eqnlabel\@vacuum}  }
\def\thebibliography#1{
\vskip 0.5cm \centerline{\bf \Large References}
\list{
[\arabic{enumi}]}{\settowidth\labelwidth{[#1]}
\leftmargin\labelwidth
\advance\leftmargin\labelsep
\usecounter{enumi}}
\def\newblock{\hskip .11em plus .33em minus .07em}
\sloppy\clubpenalty4000\widowpenalty4000
\sfcode`\.=1000\relax}

\renewcommand{\theequation}{\arabic{section}.\arabic{equation}}
\renewcommand{\section}{\setcounter{equation}{0}\@startsection
{section}{1}{0mm}{-\baselineskip}{0.5\baselineskip} {\normalfont\Large\bfseries}}
\renewcommand{\subsection}{\@startsection
{subsection}{2}{0mm}{-\baselineskip}{0.5\baselineskip} {\normalfont\large\bfseries}}
\renewcommand{\subsubsection}{\@startsection
{subsubsection}{3}{0mm}{-\baselineskip}{0.5\baselineskip}
{\normalfont\normalsize\slshape}}

\begin{document}


\begin{titlepage}
\begin{centering}
{\bm\bf \Large Gauge fixing and field redefinitions of the \\ \vspace{0.2cm}Hartle--Hawking wavefunction path integral\footnote{Based on a talk given at  ``Beyond Standard Model: From Theory to Experiments'' (BSM-2021), 29 March -- 2 April 2021, online.}}

\vspace{6mm}

 {\bf Herv\'e Partouche,$^1$\footnote{herve.partouche@polytechnique.edu} Nicolaos Toumbas$^2$\footnote{nick@ucy.ac.cy} and Balthazar de Vaulchier$^1$\footnote{balthazar.devaulchier@polytechnique.edu}}

 \vspace{3mm}

$^1$  {\em CPHT, CNRS, Ecole polytechnique, IP Paris, \\F-91128 Palaiseau, France}

$^2$ {\em Department of Physics, University of Cyprus, \\Nicosia 1678, Cyprus}

\end{centering}
\vspace{0.5cm}
$~$\\
\centerline{\bf\Large Abstract}\\
\vspace{-0.6cm}

\begin{quote}

We review some recent results concerning the Hartle--Hawking wavefunction 
of the universe. We focus on pure Einstein theory of gravity in the presence of a positive cosmological constant. We carefully implement the gauge-fixing procedure for the minisuperspace path integral, by identifying the single modulus and by using diffeomorphism-invariant measures for the ghosts and the scale factor. Field redefinitions of the scale factor yield 
different prescriptions for computing the no-boundary ground-state wavefunction. They give rise to an infinite set of ground-state wavefunctions, each satisfying a different Wheeler--DeWitt equation, at the semi-classical level. The differences in the form of the Wheeler--DeWitt equations can be traced to ordering ambiguities in constructing the Hamiltonian upon canonical quantization. However, the inner products of the corresponding Hilbert spaces turn out to be equivalent, at least semi-classically. Thus, the model yields universal quantum predictions.

\end{quote}

\end{titlepage}
\newpage
\setcounter{footnote}{0}
\renewcommand{\thefootnote}{\arabic{footnote}}
 \setlength{\baselineskip}{.7cm} \setlength{\parskip}{.2cm}

\setcounter{footnote}{0}
\setcounter{equation}{0}
\setcounter{section}{0}
\setcounter{subsection}{0}


\section{Introduction}
\label{1}

The inflationary universe scenario \cite{Guth,Linde0,Steinhardt} has been 
very successful in accounting for key cosmological puzzles of the hot Big 
Bang model, such as the flatness, the horizon and monopole problems, and in generating the primordial fluctuations that eventually led to the large scale structure observed today \cite{Mukhanov}. Furthermore, inflation has gained support by observational data concerning the anisotropies in the cosmic microwave background radiation. Despite the many successes however, there are still open questions regarding the UV completion of inflationary models and a lack of understanding of the initial conditions from first principles. To date there is no complete embedding of phenomenologically viable inflationary models in string theory.\footnote{See e.g. \cite{KKLMM} for seminal work towards this end.} We also lack of a definite understanding of {\it how} the Universe could have entered naturally into such an inflationary phase in the early past. It is likely that a quantum, probabilistic explanation exists, in terms of a wavefunction that favors suitable conditions to initiate inflation. See e.g. \cite{DeWitt, HH, Vilenkin1, Vilenkin2, Vilenkin3, Vilenkin4, Linde, Vilenkin5, HawkingL=0, HP} for different perspectives and discussions.   

A very appealing possibility to explore is to apply the no-boundary proposal of Hartle and Hawking \cite{HH}. In this context, the wavefunction of 
the universe is computed via a Euclidean path integral over all compact four-geometries that end on a particular spatial slice. The induced metric 
on this slice and the value of the inflaton field are fixed to be $h_{ij}$  and $\phi_0$, respectively. The four-geometries summed over should have no boundaries other than that of metric $h_{ij}$. As a result the wavefunction is expressed as a functional of $h_{ij}$ and $\phi_0$. We refer to this wavefunction as the ``ground state'' wavefunction, even though such a denomination may not be appropriate since in quantum gravity {\it all} physical quantum states associated with a closed universe are annihilated by the Hamiltonian. In fact, the Hartle--Hawking wavefunction can be interpreted as a probability amplitude to create from nothing a three-dimensional universe with metric $h_{ij}$ and inflaton field $\phi_0$ \cite{Vilenkin1, Vilenkin2, Vilenkin3, Vilenkin4, Vilenkin5}. As argued by Vilenkin \cite{Vilenkin1, Vilenkin2, Vilenkin3, Vilenkin4, Vilenkin5} and also 
by Linde \cite{Linde, LindeBook} some time ago, a suitable continuation to Euclidean time yields probability amplitudes favoring inflation.

In this work we revisit the Hartle--Hawking no-boundary proposal in the context of pure Einstein's theory of gravity with a positive cosmological constant $\Lambda >0$. Our goal is to discuss a number of issues pertaining to this path integral approach to quantum cosmology in a rather simpler setting, before delving into analyzing more complex cosmological models 
in the presence of matter (including inflationary ones). Indeed, in the minisuperspace approximation, where the universe is taken to be homogeneous and isotropic, the degrees of freedom reduce to a single scale factor depending only on time.  The issues we would like to discuss were recently 
raised in \cite{PTV}, in the context of the minisuperspace approximation, 
and concern i) the proper gauge fixing of the local symmetry group associated with time-reparametrization invariance; ii) the construction of an infinite set of ``ground state'' wavefunctions based on field redefinitions of the scale factor degree of freedom; iii) the derivation of the corresponding Wheeler--DeWitt equations \cite{DeWitt}; and finally, iv) the equivalence of these prescriptions at the semiclassical level and observable predictions. To our knowledge, these points have not been adequately addressed in the literature before. We believe they will prove to be important in properly applying the no-boundary proposal to obtain probabilities 
in cosmological, inflationary settings.\footnote{Previous work on the Hartle--Hawking wavefunction, related to our discussions but with some different results, includes \cite{Halli, Halli2, Turok1}. Further work and applications can be found in \cite{HalliHawk, Schleich:1986db, Halli-Hartle, 
Quevedo, Davidson1, Davidson2, pearls}.}

To illustrate these issues, it is convenient to interpret the minisuperspace model as a non-linear sigma model, where the Euclidean time parameterizes the base manifold, which is a line segment. The scale factor parameterizes a one-dimensional target space, which is a half line. 

The theory is invariant under time-reparametrizations of the base manifold. In section \ref{2}, we implement the gauge fixing procedure of Euclidean-time 
reparameterizations. The path integral over the lapse function reduces to 
an integral over the modulus of the base manifold, which can be identified to be the proper length of the line segment. We express the Faddeev--Popov determinant as a path integral over anticommuting ghost fields, and compute it to be a constant, independent of the modulus of the line segment. It is important to use gauge invariant measures in both the ghost and scale factor path integrals to implement the gauge-fixing properly.   

Field redefinitions of the scale factor, $a=A(q)$, amount to reparameterizations of the target space and leave the classical sigma model action invariant. At the quantum level, the path integral measures $\D a$ and $\D q$ are not equivalent in general, since they are related by a non-trivial Jacobian. Since there is no preferred choice, an infinite number of ground-state wavefunctions can be constructed, upon implementing the no-boundary proposal (based on the different measures $\D q$). In section \ref{3}, we compute the ground-state wavefunction for each choice of $\D q$, using the steepest-descent method, expanding around instanton solutions to 
quadratic order. The path integral over the fluctuations are obtained by applying the methods of \Refe{Coleman} -- see also \cite{Callan}. 

We proceed in section \ref{4} to determine the Wheeler--DeWitt equation each ground state wavefunction satisfies. Recall that there is an ambiguity in the exact form of the Wheeler--DeWitt equation, due to an ordering ambiguity of $q$ and its conjugate momentum $\pi_q$ in the quantum Hamiltonian. For each $\D q$, we resolve this ambiguity in the Wheeler--DeWitt equation by comparing with solutions via the WKB approximation. The inner product in each case is determined by imposing hermiticity of the corresponding Hamiltonian. Despite the fact that the precise form of the inner product depends on the choice $\D q$, the norms of the wavefunctions at the semiclassical level turn out to be the same, leading to universal predictions, independent of the $\D q$ prescription. For the particular model at hand, the norm of the wavefunctions turns out to be logarithmically divergent. At best, these wavefunctions can be used to discuss relative probabilities. We conclude in section \ref{5} with further discussion and perspectives. Throughout we work in Planck units, setting $M_p=\sqrt{8\pi 
G}=1$.

\section{{The ground-state wavefunction as a gauge fixed path integral}}
\label{2}
The Lorentzian theory is formulated on four-manifolds with space-like boundaries at initial and final times. 
The slices at constant time $x^0$ are taken to be compact and closed. In the minisuperspace approximation, these slices are restricted to be homogeneous and isotropic $3$-spheres. As a result, the physical degrees of freedom reduce to to a single scale factor depending on time, $a(x^0)$. The 

metric is given by
\be
\d s^2=-N(x^0)^2(\d x^0)^2+a(x^0)^2\, \d\Omega_3^2\, , 
\label{FRW}
\ee
where $N(x^0)\equiv \sqrt{g_{00}(x^0)}$ is the lapse function and $\d\Omega_3$ is the volume element of the unit 3-sphere of volume $v_3=2\pi^2$. Einstein's action, in the presence of a non-zero positive cosmological constant $\Lambda$, takes the form
\be
S=3v_3\int_{x^0_{\rm i}}^{x^0_{\rm f}} \d x^0\, N\bigg[\!-\!{a\over N^2}\Big({\d a\over \d x^0}\Big)^2+a-\lambda^2a^3\bigg]\, ,\quad \where \quad \lambda=\sqrt{\Lambda\over 3}\, .
\label{action2}
\ee
The classical equations of motion can be obtained by varying the action, keeping the scale factor at initial and final times, $x^0_{\rm i}$ and $x^0_{\rm f}$, fixed.\footnote{The boundary action cancels upon integrating 
by parts a bulk term that involves the second derivative of the scale factor, see e.g. \cite{PTV} for details.} 

Notice that the kinetic energy term of the scale factor has a negative sign compared to that of a conventional matter scalar field. This fact motivates us to consider two alternative prescriptions for the continuation to Euclidean time,   
\be
x^0=s\, i\, x^0_{\rm E} \, , ~~\quad \where ~~\quad s \in\{1,- 1\}\, ,
\label{LE}
\ee 
both of which have been advocated in the literature. Hartle and Hawking~\cite{HH} adopt the conventional prescription $s=-1$. In this case, the no-boundary wavefunctions become large as $\lambda \to 0$, and so they seem to favor a vanishing cosmological constant \cite{HawkingL=0}. On the 
other hand, Vilenkin~\cite{Vilenkin1, Vilenkin2,Vilenkin3,Vilenkin4,Vilenkin5} and Linde~\cite{Linde, LindeBook} have argued for $s=+1$, which favors conditions amenable for inflation. The Euclidean action  $S_{\rm E}=-iS$ in each case is given by
\be
S_{\rm E}[g_{00},a]=3sv_3\int_{x^0_{\rm Ei}}^{x^0_{\rm Ef}} \d x^0_{\rm 
E}\,  \sqrt{g_{00}}\,\bigg[ a\, g^{00}\Big({\d a\over \d x^0_{\rm E}}\Big)^2+V(a)\bigg]\, ,
\label{action3}
\ee
where
\be
V(a)=a-\lambda^2a^3\, .
\ee
This potential becomes negative when $\lambda a > 1$. When $s=-1$, the action can become arbitrarily large and negative due to rapidly oscillating configurations of the scale factor. On the other hand, for $s=1$, there are  time-independent configurations, satisfying $\lambda a \gg 1$, that yield arbitrarily large negative values for the action. We see that both choices yield Euclidean actions, which are not bounded from below, and thus a suitable continuation will be needed to obtain convergent path integrals.

Based on the form of the action $S_{\rm E}$, we interpret the theory as a 
non-linear $\sigma$-model. The base manifold is a line segment of metric $g_{00}$, parameterized by the Euclidean time $x^0$. The one-dimensional target space is parameterized by the scale factor $a$. The metric is given by
\be
G_{aa}=6v_3a\, . 
\ee
The local symmetry group consists of Euclidean-time diffeomorphisms of the base manifold. Under such a coordinate change, the metric $g_{00}$ transforms as a tensor and the scale factor as a scalar:
\be
\xi(x^0_{\rm E})=x^{\xi 0}_{\rm E}\, , ~~\quad g^\xi_{00}(x^{\xi0}_{\rm 
E}) = \bigg({\d x^0_{\rm E}\over \d x^{\xi 0}_{\rm E}}\bigg)^2\,g_{00}(x^0_{\rm E})\, , ~~\quad a^\xi(x^{\xi 0}_{\rm E})=a(x^0_{\rm E})\, .
\label{transfo}
\ee 
In addition, the action is invariant under field redefinitions of the scale factor, $a=A(q)$, which can be interpreted as reparameterizations of 
the target space.

In order to implement the no-boundary proposal, we take the initial boundary $3$-sphere to have vanishing radius, \mbox{$a(x^0_{\rm Ei})\equiv a_{\rm i}=0$}, and fix the radius of the final sphere to an arbitrary value: $a(x^0_{\rm Ef})\equiv a_{\rm f}=a_0$. We then define the ground-state wavefunction to be given by the following Euclidean path integral \cite{HH}  
\be
\Psi(a_0)=\int {\D g_{00} \over \Vol(\Diff[g_{00}])} \int_{\textstyle{\,a_{\rm i}=0, \,\, a_{\rm f}=a_0}}\,\,\D a \, e^{-{1\over \hbar}S_{\rm E}[g_{00},a]}\, , 
\label{Psi(a0)}
\ee
where we kept explicit the reduced Planck constant $\hbar$. According to Vilenkin, the wavefunction thus defined can be interpreted as the probability amplitude for creating a $3$-dimensional spherical universe of radius $a_0$ from nothing~\cite{Vilenkin1, Vilenkin2,Vilenkin3,Vilenkin4}. Notice that we have divided the measure $\D g_{00}$ in the path integral by the volume of the local symmetry group, $\Vol(\Diff[g_{00}])$, in order to take care of the overcounting of physical configurations, yielded by diffeomorphism-equivalent metrics $g_{00}$. The measure $\D a$ must be invariant under Euclidean-time diffeomorphisms. Such a gauge-invariant measure, however, is far from being unique. As we will see later on, field redefinitions of the scale factor provide us with an infinite set of inequivalent diffeomorphism-invariant measures, $\D q$, leading to an infinite set of alternative definitions for the ground state wavefunction.

We proceed now to discuss the gauge-fixing procedure, which allows us to express the wavefunction Eq. (\ref{Psi(a0)}) as an integral over physically distinct configurations. First notice that not all metrics $g_{00}$ are diffeomorphism-equivalent, since the proper length $\ell$ of the line segment remains invariant under such transformations
\be
\ell = \int_{x_{\rm Ei}}^{x_{\rm Ef}} \d x^0_{\rm E} \,\sqrt{g_{00}}=\int_{\xi(x_{\rm Ei})}^{\xi(x_{\rm Ef})} \d x^{\xi0}_{\rm E} \,\sqrt{g^\xi_{00}}\, .
\label{di}
\ee  
Thus, the proper length $\ell$ behaves as a modulus, and its value can be 
used to distinguish the classes of diffeomorphism-equivalent metrics. In \cite{PTV} we show that the line segment has no other moduli than the proper length $\ell$. Let $\hat g_{00}[1]$, defined on a domain $[\hat x^0_{\rm Ei}, \hat x^0_{\rm Ef}]$, be a fiducial metric representing the class 
$\ell=1$. Then all the other equivalence classes can be represented by fiducial metrics of the form $\hat g_{00}[\ell]=\ell^2\hat g_{00}[1]$, defined on the same interval $[\hat x^0_{\rm Ei}, \hat x^0_{\rm Ef}]$.\footnote{The Killing group of metric isometries reduces to a discrete $\Z_2$ group, generated by the transformation that reverses the orientation of 
the line segment.} 

Choosing such a metric $\hat g_{00}[\ell]$ for each equivalence class, we 
insert in Eq. (\ref{Psi(a0)}) a gauge fixing condition   
\be
1 = \Delta_{\rm FP}[g_{00}] \int_0^{+\infty}\!\!\d\ell\int_{\Diff[\hat g_{00}[\ell]]}\!\!\D\xi\; \delta\big[g_{00}-\hat g^\xi_{00}[\ell]\big]\, ,
\label{fp}
\ee
where $\Delta_{\rm FP}[g_{00}]$ is the Faddeev--Popov determinant, which is gauge invariant. Then integrating over $g_{00}$ fixes the metric to be 
$\hat g^\xi_{00}[\ell]$ (defined on $[\xi(\hat x^0_{\rm Ei}), \xi(\hat x^0_{\rm Ef})]$), as implied by the Dirac $\delta$-functional, while integrating over the gauge orbits, together with gauge invariance, lead to the cancellation of the volume of the local symmetry group $\Vol(\Diff[g_{00}])$. The wavefunction simplifies as follows
\be
\Psi(a_0)=\int_0^{+\infty}\!\!\d\ell \; \Delta_{\rm FP}[\hat g_{00}[\ell]] \int_{a(\hat x^{0}_{\rm E i})=0,\, \,\,a(\hat x^{0}_{\rm E f})=a_0}\,\D a  \, e^{-{1\over \hbar}S_{\rm E}[\hat g_{00}[\ell],a]}\, ,
\ee
where the integral over the modulus $\ell$ is an ordinary integral.

The Faddeev--Popov determinant appearing in the expression above can be related to a path integral over the diffeomorphisms that are connected to the identity as follows\footnote{Due to the fact that the orientation reversal is the only Killing isometry, the path integral over all diffeomorphisms is twice the contribution of the diffeomorphisms connected to the identity.}
\be
{1\over \Delta_{\rm FP}[\hat g_{00}[\ell]]} = 2\int_0^{+\infty}\,\,\d\ell'\int_{\Diff[\hat g_{00}[\ell']]_{\Id}}\,\,\D\xi\, \delta\big[\hat g_{00}[\ell]-\hat g^\xi_{00}[\ell']\big]\, .
\label{defD}
\ee
To compute it, we first examine the total variation of the metric $\hat g_{00}[\ell]$ under infinitesimal diffeomorphisms in the vicinity of the identity and small changes of the modulus field, 
\be
\delta \hat g_{00}[\ell]\equiv \hat g^{\Id+\delta\xi}_{00}[\ell+\delta\ell]-\hat g_{00}[\ell]=-2\hat \nabla_0\delta x_{\rm E0}+2\hat g_{00}[\ell]\,{\delta\ell\over \ell}+\cdots\, ,
\ee      
where $\hat \nabla$ is the covariant derivative with respect to $\hat g_{00}[\ell]$. Then we introduce anticommuting ghost fields. Two such fields 
are needed, $c_0$ corresponding to $\delta x_{\rm E0}$ and $b^{00}$ corresponding to the tensor field $\beta^{00}$ needed to express the $\delta$-functional as a Fourier integral \cite{PTV}. Moreover, one introduces an anticommuting variable $\lambda$ corresponding to $\delta\ell$. Berezin integration over $\lambda$ yields the following path integral expression \cite{PTV}
\be
\Delta_{\rm FP}[\hat g_{00}[\ell]]=2i\pi \alpha  \int_{c^0(\hat x^0_{\rm Ei})=0, \,\, c^0(\hat x^0_{\rm Ef})=0}   \D c\int\D b \, \Big(b,{\hat g[\ell]\over \ell}\Big)_\ell\,\exp\Big\{4i\pi \,(b,\hat \nabla c)_\ell\Big\}\, ,
\ee 
where $\alpha$ is an irrelevant constant and the tensor inner product is given by 
\be
(f, h)_{\ell}\equiv\dis  \int_{\hat x^0_{\rm Ei}}^{\hat x^0_{\rm Ef}} \d \hat x^0_{\rm E}\, \sqrt{\hat g_{00}[\ell]}\; f^{00}\, h_{00}\, . 
\ee

The ghost path integrals can be readily computed by expanding the ghost fields in Fourier modes on the interval $[\hat x^0_{\rm Ei},\hat x^0_{\rm Ef}]$. To achieve this, we must take into account the boundary conditions 
and use gauge invariant measures -- we refer the reader to \cite{PTV} for 
the detailed computations. The Faddeev--Popov determinant turns out to be 
a constant, independent of the modulus $\ell$. This is to be contrasted with the case of a base manifold with the topology of a circle, where the Faddeev--Popov  determinant is non-trivial, being proportional to $1/\ell$, where $\ell$ is the proper length of the circle. As a result the wavefunction further simplifies to the following gauge-fixed path-integral expression
\be
\label{Psigaugefixed}
\Psi(a_0)=\Delta_{\rm FP}\int_0^{+\infty}\!\!\d\ell \;  \int_{a(\hat x^{0}_{\rm E i})=0,\, \,\,a(\hat x^{0}_{\rm E f})=a_0}\,\D a  \, e^{-{1\over \hbar}S_{\rm E}[\hat g_{00}[\ell],a]}\, ,
\ee
where $\Delta_{\rm FP}$ is an irrelevant constant.

\section{{Scale factor path integral and field redefinitions}}
\label{3}
Next we compute the path integral over the scale factor and the integral over the modulus $\ell$. Since the path-integral expression (\ref{Psigaugefixed}) for the wavefunction is gauge invariant, we choose to work in a convenient gauge, setting the lapse function to be a constant,
\be
\hat g_{00}[\ell](\tau)=\ell^2\quad  \mbox{defined on} \quad [\hat x^0_{\rm Ei} ,\hat x^0_{\rm Ef} ]= [0,1]\, .
\label{lapseC}
\ee
The Euclidean-time coordinate $\hat x^0_{\rm E}$ is denoted by $\tau$. This time variable is proportional to the ``cosmological Euclidean time $t_{\rm E}$,'' which satisfies $\d t_{\rm E}=\ell\, \d\tau$. The wavefunction becomes
\be
\Psi(a_0)=\Delta_{\rm FP} \int_0^{+\infty}\,\d\ell   \int_{a(0)=0, \, 
\,a(1)=a_0}\,\D a  \, e^{-{1\over \hbar}S_{\rm E}[\ell^2,a]}\, ,
\label{psa0}
\ee
with the action (\ref{action3}) written as
\be
S_{\rm E}[\ell^2,a]=3sv_3\int_0^1\d \tau \,\bigg[ {a\over \ell} \Big({\d a\over \d \tau}\Big)^2+\ell\, V(a)\bigg]\, .
\label{SE}
\ee 
This action is not quadratic, and so we will approximate the path integral via the method of steepest-descent. To this end, we first expand the action around its extrema to quadratic order, and then carry out the resulting Gaussian integrals over the fluctuations. This steepest-descent approximation becomes accurate in the semiclassical limit, where $\hbar\to 0$.

Let us denote an extremum of the action by $(\bell^2,\ba)$, where we require the solution $\ba$ to satisfy the boundary conditions $\ba(0)=0$ and $\ba(1)=a_0$. Varying with respect to the modulus $\ell$ gives
\be
0=\!\left.{\d S_{\rm E}\over \d \ell}\right|_{(\bell^2,\ba)}=3sv_3\int_0^1\d \tau \,\bigg[ \!-\!{\ba\over \bell^2} \Big({\d \ba\over \d \tau}\Big)^2+ V(\ba)\bigg]\, ,
\label{e1}
\ee   
while the equation of motion of the scale factor can be integrated to be
\be
- {\ba\over \bell^2} \Big({\d \ba\over \d \tau}\Big)^2+ V(\ba) = {\E\over 3v_3}\, , 
\ee 
where $\E$ is an arbitrary integration constant. Eq. (\ref{e1}) implies $0=s\E$, and so it suffices to solve the Friedmann equation in order to determine the extrema of the action.

It is useful to write the Friedmann equation in the form
\be
\left({\d(\lambda \ba)\over \d(\lambda \bell \tau)}\right)^2+(\lambda \ba)^2=1\, ,
\label{eqE}
\ee 
with solution $\lambda\ba(\tau)=\pm \sin(\lambda \bell \tau+\cst)$. The 
boundary conditions $\ba(0)=0$, $\ba(1)=a_0$ set the constant to be zero and  fix the modulus $\bell$. In this work and in \cite{PTV}, we consider the case 
\be
0<\lambda a_0 <1\, , 
\ee 
leaving the case $\lambda a_0>1$ for future work. Then, there are two real instanton solutions
\bea
\lambda \ba_\epsilon (\tau)\!\!\!&=&\!\!\!\sin(\lambda  \bell_\epsilon\tau)\, , \quad \epsilon\in\{+1,-1\}\, , \nonumber\\
\where\quad~~ \lambda \bell_+ \!\!\!&=&\!\!\! \arcsin(\lambda a_0)\, , \quad\lambda \bell_- = \pi- \arcsin(\lambda a_0)\, ,
\label{aepsilon}
\eea
corresponding to parts of a $4$-sphere of radius $1/\lambda$. The $\epsilon=+1$ solution describes a cap smaller than a hemisphere, while the $\epsilon=-1$ solution describes a cap bigger than a hemisphere. The instanton actions are given by 
\be
\bar S_{\rm E}^\epsilon = s\, {2v_3\over \lambda^2}\Big[1-\epsilon \big(1-(\lambda a_0)^2\big)^{3\over 2}\Big]\, .
\label{acInst}
\ee

We now proceed to expand the action around the extremal solutions. We set 

\be
\ell=\bell_\epsilon+\delta \ell\, ,  ~~\quad a(\tau)=\ba_\epsilon (\tau)+\delta a(\tau)\, , 
\ee
where the fluctuation $\delta a(\tau)$ satisfies the boundary conditions $\delta a(0)=0$ and $\delta a(1)=0$. Thus, $\delta a(\tau)$ is an 
element of the Hilbert space of square integrable real functions on $[0,1]$, vanishing at the boundary points. This space of functions is equipped 
with the inner product  
\be
(\delta a_1,\delta a_2)_{\bell_\epsilon}=\int_0^1\d\tau\bell_\epsilon\, 
\delta a_1\, \delta a_2\, .
\label{inprod}
\ee
As a result, $\delta a(\tau)$ can be expanded in terms of the orthonormal 
sine Fourier series 
\be
\big\{\sqrt{2/\bell_\epsilon}\sin(k\pi\tau), k\in\natural^*\big\}\, . 
\ee

To quadratic order in the fluctuations, the action (\ref{SE}) can be written as follows
\be
S_{\rm E}[\ell^2,a]=\bar S_{\rm E}^\epsilon+3sv_3\int_0^1\d \tau \bell_\epsilon \Big[\delta a\, \S_\epsilon \delta a + 2 \,\delta a\, V_a(\ba_\epsilon)\,{\delta\ell\over \bell_\epsilon}+ {\delta\ell\over \bell_\epsilon}\, V(\ba_\epsilon)\, {\delta\ell\over \bell_\epsilon}\Big]\!+\O(\delta^3)\, .
\label{qua}
\ee
where the linear operator $\S_\epsilon$, given by 
\be
\S_\epsilon = -{\ba_\epsilon\over \bell_\epsilon^2}\, {\d^2\over \d\tau^2}-{1\over \bell_\epsilon^2}\, {\d\ba_\epsilon\over \d \tau}\, {\d\over \d\tau}-2\lambda^2\ba_\epsilon\, ,
\ee
is self-adjoint with respect to the inner product (\ref{inprod}): $(\delta a_1,\S_\epsilon\delta a_2)_{\bell_\epsilon}=(\S_\epsilon\delta a_1,\delta a_2)_{\bell_\epsilon}$. Here also, $V_a\equiv \d V/\d a$. Moreover, as will be seen later on, this operator is invertible when $0<\lambda a_0 
<1$, a fact that allows us to diagonalize the integrand in Eq. (\ref{qua}). For this purpose we set\footnote{We consider the odd periodic extension of the function $V_a(\ba_\epsilon)$ on the real line, so that both $\delta a$ and  $\delta \check a$ can be expanded in terms of the same sine Fourier series.}
\bea
\delta a\, \S_\epsilon \delta a + 2 \,\delta a\, V_a(\ba_\epsilon)\,{\delta\ell\over \bell_\epsilon}+ {\delta\ell\over \bell_\epsilon}\, V(\ba_\epsilon)\, {\delta\ell\over \bell_\epsilon}\!\!\!&=&\!\!\! \delta 
\check a\, \S_\epsilon \delta \check a +{\delta\ell\over \bell_\epsilon} \big[V(\ba_\epsilon)-V_a(\ba_\epsilon)\S_\epsilon^{-1} V_a(\ba_\epsilon)\big]{\delta\ell\over \bell_\epsilon}\, , \nonumber \\
\where \quad \delta \check a \!\!\!&=&\!\!\! \delta a+{\delta\ell\over \bell_\epsilon}\, \S_\epsilon^{-1}V_a(\ba_\epsilon)\, .
\label{atilde}
\eea

Using Eq. (\ref{atilde}) and defining
\be
  \K_\epsilon = \int_0^1\d\tau \bell_\epsilon\, \big[V(\ba_\epsilon)-V_a(\ba_\epsilon)\S_\epsilon^{-1} V_a(\ba_\epsilon)\big]\, ,
 \label{barI}
\ee
we obtain the following expression for the wavefunction (\ref{psa0}) in the steepest-descent approximation,
\bea
\Psi(a_0)\!\!\!&=&\!\!\!\Delta_{\rm FP}\sum_{\epsilon=\pm 1} e^{-{1\over \hbar} \bar S_{\rm E}^\epsilon}\,Z_\epsilon(a_0) \int \d\delta \ell \,\exp\Big\{\!-\!{3sv_3\over \hbar}\,  \K_\epsilon  \Big({\delta\ell\over \bell_\epsilon}\Big)^2\Big\}\,(1+\O(\hbar))\nonumber \\
\where \quad Z_\epsilon(a_0)\!\!\!&=&\!\!\! \int_{\delta \check 
a(0)=0, \, \,\delta \check a(1)=0}\D \delta \check a  \, \exp\Big\{\!-\!{3sv_3\over \hbar}\,( \delta \check a, \S_\epsilon \delta \check a)_{\bell_\epsilon}\Big\}\, .
\label{psii}
\eea


The operator $\S_\epsilon$ is self-adjoint and so it can be diagonalized in an orthonormal basis. Let us denote its eigenvectors by $\phi^\epsilon_k$ and the corresponding eigenvalues by $\nu_k^\epsilon$. These satisfy
\be
\S_\epsilon\phi^\epsilon_k = \nu^\epsilon_k\phi^\epsilon_k\, , ~~ k\in\natural^*\, , \quad~~\where~~\quad  (\phi^\epsilon_k,\phi^\epsilon_{k'})_{\bell_\epsilon}=\delta_{kk'}\, ,~~ \nu^\epsilon_k\in\R\, .
\ee
Then we may expand the scale factor fluctuation as
\be
\delta {\check a}(\tau)=\sum_{k\ge 1}{\bm \delta {\check a}_k}\,  \phi^\epsilon_k(\tau)\, , 
\label{expanda}
\ee
and use zeta regularization to get
\bea
Z_\epsilon(a_0)\!\!\!&=&\!\!\!\prod_{k\ge 1}\int d{\bm \delta {\check a}_k}\, e^{-{3sv_3\over \hbar} \nu^\epsilon_k({\bm \delta {\check a}_k})^2}= \prod_{k\ge 1} \sqrt{\hbar\, \pi\over 3sv_3\, \nu^\epsilon_k}\nonumber\\
\!\!\!&=& \!\!\!\left({3sv_3\over \hbar \, \pi}\right)^{1\over 4}{1\over \sqrt{\det \S_\epsilon}}\, . 
\label{Ze}
\eea
In order to define the Gaussian integrals, we have used the following prescription: The Fourier mode ${\bm {\delta {\check a}_k}}$ is integrated from $-\infty$ to $+\infty$ when $s \nu_k^\epsilon>0$, and from $-i\infty$ 
to $+i\infty$ when $s \nu_k^\epsilon<0$. There is no vanishing eigenvalue, $\nu_k^\epsilon=0$, since $\S_\epsilon$ is invertible (see below). In 
fact, $\det \S_+$ and $\det \S_-$ turn out to have opposite signs, independently of the sign of $s$. 
Hence, rotating some contours of integration along the imaginary axis is necessary for both $Z_{+}(a_0)$ and $Z_{-}(a_0)$ to exist, irrespectively 
of the choice of continuation to Euclidean time.

The determinant of $\S_\epsilon$ can be computed via the method of \Refe{Coleman}. It is given by 
\be
\det \S_\epsilon=\N_\epsilon\, \varphi^{\epsilon}_{0}(1)\, ,
\label{detfor}
\ee
where $\N_\epsilon$ is a universal constant and the function $\varphi^\epsilon_{0}(\tau)$ (to be evaluated at $\tau =1$) solves the system
\be
\left\{\begin{array}{l}
\S_\epsilon \varphi^\epsilon_0(\tau) = 0 \, , \\
\varphi^\epsilon_0(\tau_\epsilon)=0\, , \quad \dis {\d\varphi^\epsilon_0\over \d\tau}(\tau_\epsilon)=1\, .\end{array}\right.
\label{difs}
\ee
Here, $\tau_\epsilon\in(0,1)$ is a regulator to be sent to $0$ at the end 
of the calculations. The universal constant $\N_\epsilon$ can be obtained 
by finding the determinant of an operator that is identical to $\S_\epsilon$ up to terms involving no derivatives. The computations of $\N_\epsilon$ and $\varphi^{\epsilon}_{0}(1)$ have been carried out in great detail in \cite{PTV}, giving the net result
\be
\det \S_\epsilon = 2\, \Big({\theta_*\over \lambda}\Big)^{1\over 4}\ln{1\over \theta_*}\times \epsilon\, a_0^{1\over 4}\sqrt{1-(\lambda a_0)^2}\, ,
\label{final_detS}
\ee
where $\theta_*=\lambda {\bar\ell}_\epsilon\tau_\epsilon$ is to be sent 
to zero.
Note that
\be
\det\S_+>0\, ,\quad \det\S_-<0\, , \quad \when \quad 0<\lambda a_0<1\, ,
\ee
demonstrating that both $\S_+$ and $\S_-$ are invertible. 

The integral over the fluctuation $\delta \ell$ is Gaussian. To evaluate it we need to determine $ \K_\epsilon$ given in Eq. (\ref{barI}), which requires to find the function $\S_\epsilon^{-1}V_a(\ba_\epsilon)$, or equivalently the function $f_\epsilon$ satisfying $\S_\epsilon f_\epsilon = 
V_a(\ba_\epsilon)$ along with the boundary conditions $f_\epsilon(\tau_\epsilon)=f_\epsilon(1)=0$. These yield the following net result \cite{PTV} 
\be
\int \d\delta \ell \,\exp\Big\{\!-\!{3sv_3\over \hbar}\,  \K_\epsilon  \Big({\delta\ell\over \bell_\epsilon}\Big)^2\Big\}=\sqrt{{\pi\hbar\over 3sv_3}\, \ln{1\over \theta_*}}\, ,
\label{fluI}
\ee
where the domain of integration is from $-\infty$ to $+\infty$ for $s=+1$, and from $-i\infty$ to $+i\infty$ for $s=-1$. 

Collecting all results in the expression for the wavefunction, Eq. (\ref{psii}), we obtain the result
\be
\Psi(a_0) = \C_s(\theta_*)\sum_{\epsilon=\pm 1} {1\over \sqrt{\epsilon}}\, {\exp\!\Big[\epsilon s\, \dis{2v_3\over \hbar\lambda^2}\,\big(1-(\lambda a_0)^2\big)^{3\over 2}\Big]\over a_0^{1\over 8}\,\big(1-(\lambda a_0)^2\big)^{1\over 4}}\, (1+\O(\hbar))\, ,\quad 0<\lambda a_0<1\, ,
\label{psia00}
\ee
where 
\be
\C_s(\theta_*)=\alpha\,\sqrt{i\pi} \, \Big({\pi \hbar\over 3sv_3}\Big)^{1\over 4}\exp\!\Big[\!-\!s\, {2v_3\over \hbar \lambda^2}\Big]\Big({\lambda \over \theta_*}\Big)^{1\over 8}
\ee
is a regulator-dependent coefficient, which is irrelevant once $\Psi(a_0)$ is normalized or when we discuss relative probabilities.   

\paragraph{Field Redefinitions:}
Let us now discuss the issue of field redefinitions. As we have already remarked, they leave the classical action invariant. They can be thought of as reparameterizations of the target space. Let us consider such a field redefinition 
\be
a=A(q) ~~ \Longleftrightarrow ~~ q=Q(a)\, , 
\ee
where $Q=A^{-1}$ is an invertible function defined for $a> 0$. The field $q(\tau)$ satisfies the following fixed boundary conditions  
\be
q(1)\equiv q_0=Q(a_0)\, ,  \quad~~ q(0)= Q(0)\, .
\ee 
The fluctuations around the instanton solutions satisfy 
\be
\delta a=A'(\bq_\epsilon)\delta q+\O((\delta q)^2)\, ,~~\quad\where\quad ~~\bq_\epsilon=Q(\ba_\epsilon)\, , 
\label{pertur}
\ee
where a prime denotes a derivative.

At the quantum level, the path integral measures $\D a$ and $\D q$ will not be equivalent in general, since they will be related by a non-trivial Jacobian. As a result, we can define a quantum wavefunction, as in Eq. (\ref{psa0}),
\be
\widetilde\Psi(q_0)=\Delta_{\rm FP} \int_0^{+\infty}\!\!\d\ell   \int_{q(0)=Q(0),\,\,q(1)=q_0}\D q  \, e^{-{1\over \hbar}\widetilde S_{\rm E}[\ell^2,q]}\, ,
\label{psa1}
\ee
based on the gauge invariant path-integral measure $\D q$. The tilde action satisfies
\be
S_{\rm E}[\ell^2,a]\equiv \widetilde S_{\rm E}[\ell^2,q] \, .
\label{stilde}
\ee
Following similar steps as before, we may calculate $\widetilde\Psi(q_0)$ 
in the semi-classical limit to get \cite{PTV}  
\be
\wt\Psi(q_0) = {\wt\C}_s(\theta_*)\sum_{\epsilon=\pm 1} {1\over \sqrt{\epsilon\,{\rm sign}(Q')}}\, {\exp\!\Big[\epsilon s\, \dis{2v_3\over \hbar\lambda^2}\,\big(1-(\lambda a_0)^2\big)^{3\over 2}\Big]\over a_0^{1\over 8}\;|Q'(a_0)|^{1\over 4}\,\big(1-(\lambda a_0)^2\big)^{1\over 4}}\, (1+\O(\hbar))\, ,\quad 0<\lambda a_0<1\, ,
\label{1st ex}
\ee
where $\wt\C_s(\theta_*)$ is a regulator-dependant coefficient 
\be
\wt\C_s(\theta_*)=\alpha\,\sqrt{i\pi} \, \Big({\pi \hbar\over 3sv_3}\Big)^{1\over 4}\exp\!\Big[\!-\!s\, {2v_3\over \hbar \lambda^2}\Big]\Big({\lambda \over \theta_*}\Big)^{1\over 8}\, \Big|Q'\Big({\sin\theta_*\over \lambda}\Big)\Big|^{-{1\over 4}}\, .
\ee
The wavefunction can  also be expressed in terms of $q_0$. The expression 
is 
\be
\wt\Psi(q_0) = {\wt\C}_s(\theta_*)\sum_{\epsilon=\pm 1} {1\over \sqrt{\epsilon\,{\rm sign}(A')}}\, {\exp\!\Big[\epsilon s\, \dis{2v_3\over \hbar\lambda^2}\,\big(1-(\lambda A(q_0))^2\big)^{3\over 2}\Big]\over A(q_0)^{1\over 8}\;|A'(q_0)|^{-{1\over 4}}\,\big(1-(\lambda A(q_0))^2\big)^{1\over 4}}\, (1+\O(\hbar))\, .
\label{psifinal}
\ee

We conclude that there are infinitely many prescriptions to define the ``ground state'' wavefuction. In the next section we will show that these yield identical observable predictions.

\section{{ Wheeler--DeWitt equation and universality}}
\label{4}
For each choice $\D q$, the corresponding ground-state  wavefunction satisfies a Wheeler--DeWitt equation. To see this let us first note that the path integral of a total functional derivative must vanish
\be
0=\int {\D N \over \Vol(\Diff[N^2])}\, {\delta\over \delta N(x^0)}\, e^{i\wt S[N^2,q]}\, , \quad \mbox{for all $x^0$}\, .
\label{functional}
\ee
In this formula, $\wt S$ is the Lorentzian action expressed in terms of the field $q$ and  corresponding to the classical Lagrangian 
\be
\wt L(N,q,\dot q)= 3 v_3\Big(\!-\!{A(q)A'(q)^2\over N}\,\dot q^2+N \wt V(q)\Big) \, .
\ee
Using this expression, it is easy to see that Eq. (\ref{functional}) further yields the constraint identity 
\be
0=-i \int_\C {\D N \,\D q\over \Vol(\Diff[N^2])} \left.{ \wt H\over N}\right|_{x^0}\, e^{i\wt S[N^2,q]}\, ,
\ee
where 
\be
\wt H = N\Big(\!- \!{1\over 12v_3}\, {\pi_q^2\over AA'^2}-3v_3\wt V\Big)\,  
\label{H/N}
\ee
is the classical Hamiltonian. Here, $\pi_q$ is the momentum conjugate to $q$ given by
\be
\pi_q ={\partial\wt L\over \partial \dot q} = -6 v_3{AA'^2\over N}\,\dot q\, . 
\ee 
The implication of the constraint identity is the vanishing of all matrix 
elements of the quantum Hamiltonian divided by the lapse function. Equivalently, the quantum Hamiltonian (divided by the lapse function) must annihilate all physical states. The corresponding wavefunctions must satisfy the Wheeler--DeWitt equation.

As usual, the canonical quantization of the classical expression for $\wt 
H/N$ can be obtained by replacing  
\be
q\longrightarrow  q_0\, ,  ~~\quad \pi_q\longrightarrow  -i\hbar\, {\d\over \d q_0}\, , 
\ee
which satisfy the canonical commutation relation $[q,\pi_q]=i\hbar$. However, because the first term in the classical expression of $\wt H/N$ involves a product of functions $q$ and $\pi_q$, there are ordering ambiguities in constructing the quantum operator. These ambiguities are induced in the precise form of the Wheeler--DeWitt equation. They can be parameterized in terms of two functions of $q$, $\widetilde \rho$ and $\widetilde 
\omega$, as follows \cite{PTV}
\be
{\wt H\over N}\,\wt\Psi_\C\equiv {\hbar^2\over 12v_3}\,  {1\over AA'^2} \bigg[{1\over \wt \rho}\, {\d\over \d q_0}\Big(\wt \rho\, {\d\wt \Psi_\C\over \d q_0}\Big)+\wt \omega\wt\Psi_\C\bigg]- 3v_3\wt V\wt \Psi_\C=0\, ,
\label{WDWtil}
\ee
where $\Psi_\C$ denotes a generic solution. 
Setting
\be
\Psi_{A\C}(a_0)\equiv \wt\Psi_\C(Q(a_0))\, ,
\label{chvar}
\ee
we may alternatively write the above equation  in terms of the scale factor as
\bea
{\wt H\over N}\,\wt\Psi_\C\equiv{H_A\over N}\,\Psi_{A\C} \!\!\! &\equiv& \!\!\!{\hbar^2\over 12v_3}\,  {1\over a_0} \bigg[{1\over \rho_A}\, {\d\over \d a_0}\Big(\rho_A\, {\d\Psi_{A\C}\over \d a_0}\Big)+ \omega_A 
\Psi_{A\C}\bigg]- 3v_3 V  \Psi_{A\C}=0\, ,\nonumber \\
\where\quad~~ \rho_A(a_0)\!\!\!&=&\!\!\! {\wt\rho(Q(a_0))\over |Q'(a_0)|}\, , ~~\quad \omega_A(a_0) = \wt\omega(Q(a_0))\, Q'(a_0)^2\, .
\label{WDWtil2}
\eea

We can lift the ambiguity in the form of the Wheeler--DeWitt equation by imposing that the ground-state wavefunctions have to satisfy it. Indeed, the generic solutions at the semi-classical  can be obtained by applying the WKB method \cite{wkb}, which leads to \cite{PTV}
\be
\wt\Psi_\C(q_0) = \sum_{\epsilon=\pm 1} N_{\C\epsilon}\, {\exp\!\Big[\epsilon s\, \dis{2v_3\over \hbar\lambda^2}\,\big(1-(\lambda A(q_0))^2\big)^{3\over 2}\Big]\over |\wt \rho(q_0)|^{1\over 2}\,A(q_0)^{1\over 2}\,|A'(q_0)|^{1\over 2}\, \big(1-(\lambda A(q_0))^2\big)^{1\over 4}}\, (1+\O(\hbar))\, ,\quad  0<\lambda A(q_0)<1\, ,
\ee
where $N_{\C\epsilon}$ are two integration constants. Comparing with Eq.  
(\ref{psifinal}) we find  $\widetilde\rho$,
\be
\tilde \rho(q_0) = A(q_0)^{-{3\over 4}}\,  |A'(q_0)|^{-{3\over 2}}\, . 
\label{rhoo1}
\ee
Notice that the unknown function $\wt\omega$ is absorbed in $\O(\hbar)$ terms, and so it cannot be determined at the semi-classical level. 
The expression for $\rho_A$ is 
\be
\rho_A(a_0) = a_0^{-{3\over 4}}\,  |Q'(a_0)|^{1\over 2}.
\ee
Both  $\wt \rho(q_0)$ and $\rho_A(a_0)$ are positive for $0<\lambda a_0<1$. 
The values $N_\epsilon$ of the mode coefficients $N_{\C\epsilon}$ that select the corresponding ground-state wavefunction are given by
\be
N_\epsilon = {1\over \sqrt{\epsilon\,  {\rm sign}(Q')}}\, .
\ee

\paragraph{Quantum equivalence at the semi-classical level:}
A natural question that arises is whether different wavefunction prescriptions based on the path integral measures $\D q$, and the corresponding Wheeler--DeWitt equations, define different quantum gravity models with same classical limits. The answer to this question is negative. The reason is that all these prescriptions yield the same observable predictions at the semi-classical level.

Indeed to obtain probability amplitudes, we need to define a suitable inner product in each Hilbert space. This takes the form
\be
\langle \Psi_{A1},\Psi_{A2}\rangle_A =\int_0^{+\infty} \d a_0\,  \mu_A(a_0) \,  \Psi_{A1}(a_0)^*\,  \Psi_{A2}(a_0)\, ,
\label{pr1}
\ee
for some real positive measure $\mu_A$. 

Based on the form of the inner product, we obtain the following identity
\be
\big\langle \Psi_{A1},{H_A\over N}\,\Psi_{A2}\big\rangle = \big\langle {H_A^\dag\over N}\,\Psi_{A1},\Psi_{A2}\big\rangle+ {\hbar^2\over 12 v_3}\left[\rho_A\!\left({\mu_A\over a_0\rho_A}\, \Psi^*_{A1} \, {\d \Psi_{A2}\over \d a_0}-{\d\over \d a_0}\Big({\mu_A\over a_0\rho_A}\, \Psi^*_{A1}\Big)\Psi_{A2}\right)\right]_0^{+\infty}\, , 
\label{con}
\ee
where integration by parts gives
\be
{H^\dag_A\over N}\,\Psi_{A\C}\equiv {\hbar^2\over 12v_3}\,  {1\over a_0}\bigg[{a_0\over \mu_A}\, {\d\over \d a_0}\Big(\rho_A\, {\d\over \d a_0}\Big({\mu_A\over a_0\rho_A}\, \Psi_{A\C}\Big)+ \omega_A \Psi_{A\C}\bigg]- 3v_3 V  \Psi_{A\C}\, .
\label{hdag}
\ee
Imposing hermiticity of the Hamiltonian gives rise to a differential equation, which determines the measure $\mu_A$ in terms of $\rho_A$ \cite{PTV}:
\be
\mu_A (a_0)= a_0\, \rho_A(a_0)\, . 
\label{fin}
\ee 
Furthermore the Wheeler--DeWitt equation ensures the vanishing of the boundary term in Eq. (\ref{con}).

It follows that at the semi-classical level, the probability amplitudes $\sqrt{\mu_A}\,\Psi_{A\C}$ are universal, since 
\be
\sqrt{\mu_A(a_0)} \Psi_{A\C}(a_0) =  \sum_{\epsilon=\pm 1} N_{\C\epsilon}\, {\exp\!\Big[\dis \epsilon s\, {2v_3\over \hbar\lambda^2}\,\big(1-(\lambda a_0)^2\big)^{3\over 2}\Big]\over \dis \big(1-(\lambda a_0)^2\big)^{1\over 4}}\, (1+\O(\hbar))\,, \quad 0<\lambda a_0<1\, .
\ee
This universality relation can be extended also for $\lambda a_0 >1$ \cite{PTV}. So all probabilities and relative probabilities are independent of the choice of the path integral measure $\D q$, at least at the semi-classical level. 

An important consequence however is that none of the solutions of the Wheeler--DeWitt equation is normalizable. Indeed by examining the large $a_0$ behavior of these functions, we can infer that $|\sqrt{\mu_A(a_0)} \Psi_{A\C}(a_0)|^2$ scales as $1/a_0$ in this limit, giving rise to a logarithmically divergent norm. So at best we can use these wavefunctions to define relative probabilities, in terms of ratios of the probability densities evaluated at different points of minisuperspace in this model. It would be interesting to extend the analysis to more realistic cases, in the presence of matter, in order to see if normalizable wavefunctions, based on the no-boundary proposal, can be constructed. Interesting attempts to extract observables in quantum cosmology includes Refs.~\cite{Vilenkin5, Gibbons, HHH1, HHH2, HHH3}.

\section{{Conclusions}}
\label{5}
In this work we have considered the Hartle--Hawking wavefunction for spatially closed universes, with positive cosmological constant $\Lambda>0$. We focused on the simpler minisuperspace version, considering homogeneous 
and isotropic universes. The system can be seen as a non-linear $\sigma$-model with a line segment for the base and a one-dimensional target space 
parameterized by the scale factor. The gauge fixing of time reparameterizations is achieved by integrating over the proper length of the line-segment base, introducing the necessary Faddeev--Popov determinant, which turns out to be trivial, and using gauge invariant measures for the scale factor path integral. The reparametrizations of the scale factor, that is the coordinate of the target space, yield different gauge invariant measures and path integrals, but the corresponding Hilbert spaces are equivalent, at least semi-classically. 


\bibliographystyle{unsrt}

\begin{thebibliography}{99}
\bibitem{Guth} A.~H.~Guth,
``The inflationary universe: A possible solution to the horizon and flatness problems,''
Phys. Rev. D \textbf{23}, 347-356 (1981).

\bibitem{Linde0} 
A.~D.~Linde,
``A new inflationary universe scenario: A possible solution of the horizon, flatness, homogeneity, isotropy and primordial monopole problems,''
Phys. Lett. B \textbf{108}, 389-393 (1982).

\bibitem{Steinhardt} 
A.~Albrecht and P.~J.~Steinhardt,
``Cosmology for grand unified theories with radiatively induced symmetry breaking,''
Phys. Rev. Lett. \textbf{48}, 1220-1223 (1982).

\bibitem{Mukhanov}
V.~F.~Mukhanov and G.~V.~Chibisov,
``Quantum fluctuations and a nonsingular universe,''
JETP Lett. \textbf{33}, 532-535 (1981);
V.~F.~Mukhanov, H.~A.~Feldman and R.~H.~Brandenberger,
``Theory of cosmological perturbations. Part 1. Classical perturbations. Part 2. Quantum theory of perturbations. Part 3. Extensions,''
Phys. Rept. \textbf{215}, 203-333 (1992).

\bibitem{KKLMM}
S.~Kachru, R.~Kallosh, A.~D.~Linde, J.~M.~Maldacena, L.~P.~McAllister and 
S.~P.~Trivedi,
``Towards inflation in string theory,''
JCAP \textbf{10}, 013 (2003)
[arXiv:hep-th/0308055 [hep-th]].

\bibitem{DeWitt}
B.~S.~DeWitt,
``Quantum theory of gravity. I. The canonical theory,''
Phys. Rev. \textbf{160} (1967), 1113-1148.

\bibitem{HH} 
  J.~B.~Hartle and S.~W.~Hawking,
  ``Wave function of the universe,''
  Phys.\ Rev.\ D {\bf 28} (1983) 2960
   [Adv.\ Ser.\ Astrophys.\ Cosmol.\  {\bf 3} (1987) 174].

 \bibitem{Vilenkin1}
  A.~Vilenkin,
  ``Creation of universes from nothing,''
  Phys.\ Lett.\ B {\bf 117} (1982) 25.
  
\bibitem{Vilenkin2}
  A.~Vilenkin,
  ``The birth of inflationary universes,''
  Phys.\ Rev.\ D {\bf 27} (1983) 2848.
  
\bibitem{Vilenkin3} 
  A.~Vilenkin,
  ``Quantum creation of universes,''
Phys. Rev. D \textbf{30} (1984), 509.
    
\bibitem{Vilenkin4}
A.~Vilenkin,
``Boundary conditions in quantum cosmology,''
Phys. Rev. D \textbf{33} (1986), 3560.

\bibitem{Linde} 
  A.~D.~Linde,
  ``Quantum creation of an inflationary universe,''
  Sov.\ Phys.\ JETP {\bf 60}, 211 (1984)
  [Zh.\ Eksp.\ Teor.\ Fiz.\  {\bf 87}, 369 (1984)]; A.~D.~Linde,
  ``Quantum creation of the inflationary universe,''
  Lett.\ Nuovo Cim.\  {\bf 39}, 401 (1984).

\bibitem{Vilenkin5}
A.~Vilenkin,
``Predictions from quantum cosmology,''
NATO Sci. Ser. C \textbf{476} (1996), 345-367
[arXiv:gr-qc/9507018 [gr-qc]]. 

\bibitem{HawkingL=0}
S.~W.~Hawking,
``The cosmological constant is probably zero,''
Phys. Lett. B \textbf{134} (1984), 403.

\bibitem{HP}
S.~W.~Hawking and D.~N.~Page,
``How probable is inflation?,''
Nucl. Phys. B \textbf{298}, 789-809 (1988).


\bibitem{LindeBook}
A.~D.~Linde,
``Particle physics and inflationary cosmology,''
Contemp. Concepts Phys.~\textbf{5} (1990), 1-362
[arXiv:hep-th/0503203 [hep-th]].

\bibitem{PTV}
H.~Partouche, N.~Toumbas and B.~de Vaulchier,
``Wavefunction of the universe: Reparametrization invariance and field redefinitions of the minisuperspace path integral,''
[arXiv:2103.15168 [hep-th]].

\bibitem{Halli}
J.~J.~Halliwell,
``Derivation of the Wheeler--DeWitt equation from a path integral for minisuperspace models,''
Phys. Rev. D \textbf{38} (1988), 2468.

\bibitem{Halli2}
J.~J.~Halliwell and J.~Louko,
``Steepest descent contours in the path integral approach to quantum cosmology. 1. The de Sitter minisuperspace model,''
Phys. Rev. D \textbf{39} (1989), 2206.

\bibitem{Turok1}
J.~Feldbrugge, J.~L.~Lehners and N.~Turok,
``Lorentzian quantum cosmology,''
Phys. Rev. D \textbf{95} (2017) no.10, 103508
[arXiv:1703.02076 [hep-th]].

\bibitem{HalliHawk}
J.~J.~Halliwell and S.~W.~Hawking,
``The origin of structure in the universe,''
Phys. Rev. D \textbf{31} (1985), 1777.

\bibitem{Schleich:1986db}
K.~Schleich,
``Semiclassical wave function of the universe at small three geometries,''
Phys. Rev. D \textbf{32} (1985), 1889-1898.

\bibitem{Halli-Hartle}
J.~Diaz Dorronsoro, J.~J.~Halliwell, J.~B.~Hartle, T.~Hertog and O.~Janssen,
``The real no-boundary wave function in Lorentzian quantum cosmology,''
Phys. Rev. D \textbf{96} (2017) no.4, 043505
[arXiv:1705.05340 [gr-qc]].

\bibitem{Quevedo}
S.~Cespedes, S.~P.~de Alwis, F.~Muia and F.~Quevedo,
``Lorentzian vacuum transitions: Open or closed universes?,''
[arXiv:2011.13936 [hep-th]].

\bibitem{Davidson1}
A.~Davidson, D.~Karasik and Y.~Lederer,
``Wavefunction of a brane-like universe,''
Class. Quant. Grav. \textbf{16} (1999), 1349-1356
[arXiv:gr-qc/9901003 [gr-qc]].

\bibitem{Davidson2}
A.~Davidson and B.~Yellin,
``Quantum black hole wave packet: Average area entropy and temperature dependent width,''
Phys. Lett. B \textbf{736} (2014), 267-271
[arXiv:1404.5729 [gr-qc]].

\bibitem{pearls}
J.~J.~Halliwell and R.~C.~Myers,
``Multiple sphere configurations in the path integral representation of the wave function of the universe,''
Phys. Rev. D \textbf{40} (1989), 4011.

\bibitem{Coleman}
S.~Coleman,
``Aspects of symmetry: Selected Erice lectures,''
Cambridge University Press (2010).

\bibitem{Callan}
C.~G.~Callan, Jr. and S.~R.~Coleman,
``The fate of the false vacuum. II. First quantum corrections,''
Phys. Rev. D \textbf{16} (1977), 1762-1768.

\bibitem{wkb}
See \eg D.~J.~Griffiths,
``Introduction to quantum mechanics,'' Cambridge University Press (2016). 


\bibitem{Gibbons}
G.~W.~Gibbons, S.~W.~Hawking and J.~M.~Stewart,
``A natural measure on the set of all universes,''
Nucl. Phys. B \textbf{281} (1987), 736.


\bibitem{HHH1}
J.~B.~Hartle, S.~W.~Hawking and T.~Hertog,
``Quantum probabilities for inflation from holography,''
JCAP \textbf{01} (2014), 015
[arXiv:1207.6653 [hep-th]].

\bibitem{HHH2}
J.~Hartle, S.~W.~Hawking and T.~Hertog,
``Local observation in eternal inflation,''
Phys. Rev. Lett. \textbf{106} (2011), 141302
[arXiv:1009.2525 [hep-th]].

\bibitem{HHH3}
J.~B.~Hartle, S.~W.~Hawking and T.~Hertog,
``No-boundary measure of the universe,''
Phys. Rev. Lett. \textbf{100} (2008), 201301
[arXiv:0711.4630 [hep-th]].

 
\end{thebibliography}

\end{document}